\documentclass[10pt,twocolumn]{article}
\usepackage[T1]{fontenc}

\usepackage[a4paper,margin=1.8cm]{geometry}

\usepackage{amsmath}
\usepackage{amssymb}
\usepackage{siunitx}
\DeclareSIUnit\arbunit{arb.~units}

\usepackage{graphicx}
\usepackage{overpic}
\usepackage{colortbl}
\usepackage{xcolor}
\usepackage{dblfloatfix}
\usepackage{float}

\usepackage{anyfontsize}
\usepackage{steinmetz}
\usepackage{authblk}

\usepackage{tabularx}
\usepackage{array}

\graphicspath{{./images/}}

\usepackage[
    colorlinks=true,
    linkcolor=blue,
    citecolor=blue,
    urlcolor=blue
]{hyperref}

\title{Deterministically Fabricated C-Band Single-Photon Sources Based on Quantum Dot Circular Bragg Grating Cavities}

\author[1]{Monica Pengerla}
\author[2,3]{Yury Berdnikov}
\author[1]{Alexander Kosarev}
\author[1]{Dominik M. Feder}
\author[1]{Avijit Barua}
\author[1]{Aris Koulas-Simos}
\author[2,3,4]{Paweł Holewa}
\author[1]{Sven Rodt}
\author[1]{Imad Limame\thanks{Corresponding author: \texttt{imad.limame@tu-berlin.de}}}
\author[2,3]{Elizaveta Semenova}
\author[1]{Stephan Reitzenstein}

\affil[1]{Institute of Physics and Astronomy, Technical University of Berlin, 10623 Berlin, Germany}
\affil[2]{DTU Electro, Technical University of Denmark, 2800 Kongens Lyngby, Denmark}
\affil[3]{NanoPhoton -- Center for Nanophotonics, Technical University of Denmark, 2800 Kongens Lyngby, Denmark}
\affil[4]{Department of Experimental Physics, Faculty of Fundamental Problems of Technology, Wrocław University of Science and Technology, 50-370 Wrocław, Poland}

\affil[*]{\texttt{imad.limame@tu-berlin.de}}

\date{}

\begin{document}

\twocolumn[
\maketitle

\vspace{-1em}

\begin{center}
\end{center}

\vspace{0.5em}

\begin{abstract}
Telecom-wavelength quantum light sources are essential components for scalable fiber-based quantum networks, where on-demand emission of single photons is required for high secret key rates. Here, we demonstrate the deterministic integration of metal-organic vapor-phase epitaxy (MOVPE)-grown C-band InAs/InP quantum dots into circular Bragg grating resonators. The integrated emitters exhibit a radiative lifetime of $T_1 = (1400 \pm 50$)~ps and a spectral linewidth of $\Delta E_{\mathrm{FWHM}} = (41.0 \pm 4.0)~\mu\mathrm{eV}$. Under non-resonant excitation, we measure a post-selected single-photon purity of \((0.95 \pm 0.02)\), a raw two-photon interference visibility of \(V_{12\mathrm{ns}} = (35 \pm 2)\%\) which increases up to $(70 \pm 18)\%$ under temporal post-selection. We discuss strategies to improve photon indistinguishability, including radiative-rate engineering, charge stabilization, and more selective excitation. Our results combine MOVPE growth, deterministic device integration, and telecom C-band emission, providing a route toward scalable semiconductor quantum light sources for fiber-based quantum networks.
\end{abstract}

\vspace{1em}
]
\section{Introduction}\label{sec1}

Solid-state quantum dots (QDs) are a leading platform for generating nonclassical light for photonic quantum information technologies. They can emit single photons on demand with high purity and indistinguishability and, in suitably engineered systems, produce entangled photon pairs \cite{Heindel2023}. Their compatibility with established semiconductor growth and nanofabrication processes offers a route to scalable quantum photonic devices. Moreover, integration of QDs with low-loss dielectric waveguides and photonic integrated circuits can be used to create on-chip quantum photonic circuits \cite{Schnauber2019,Uppu2020,Wang2025,Nielsen2025}. Importantly, the QD emission wavelength can be engineered during epitaxial growth, while post-growth processing and device integration provide further control over properties such as fine-structure splitting (FSS), radiative lifetime, and coupling to the photonic environment \cite{MartnSnchez2017,Hfer2019,Gerhardt2020,Wijitpatima2024,Phillips2024}.

Telecom-wavelength operation is important for fiber-based quantum communication and distributed quantum information processing \cite{Uppu2021,Vajner2022,Yu2023}. The O- and C-bands, centered at approximately 1310 and 1550~nm, respectively, correspond to low-loss transmission windows in silica optical fibers, which underpin modern telecommunication networks. Emitters that operate directly in these bands can use existing fiber infrastructure and reduce the need for frequency conversion, thereby avoiding associated sources of loss and noise. Telecom-band QDs therefore offer a promising route to linking deterministic quantum light sources with deployed fiber networks and scalable quantum-photonic architectures~\cite{holewa_solid-state_2025}.

Despite this promise, telecom-band QDs have historically exhibited optical performance that lags behind their shorter-wavelength counterparts \cite{Yu2023}. In particular, achieving simultaneously high single-photon purity, long optical coherence, and high photon indistinguishability remains challenging despite recent progress for both O-band \cite{Albrechtsen2026} and C-band~\cite{Hauser2026} QD-SPSs. In fact, extending the emission wavelength toward the telecom regime requires changes in the QD composition and surrounding semiconductor environment that can introduce additional sources of disorder, charge fluctuations, and non-radiative recombination. These effects can manifest as spectral diffusion, dephasing, temporal jitter, and fluctuations of the local electrostatic environment, all of which directly degrade the coherence of the emitted photons \cite{Yu2023}.

The challenge of achieving high photon indistinguishability becomes particularly relevant under non-resonant excitation \cite{Hauser2026}. Above-band excitation provides a technologically attractive and experimentally robust means of generating carriers, but it also creates a large reservoir of free carriers in the semiconductor matrix surrounding the QDs. Carrier relaxation, trapping, charge fluctuations, and subsequent recapture by the QD can introduce temporal and spectral correlations between successive emission events. As a result, obtaining high-quality single-photon emission often requires increasingly sophisticated excitation and photonic-engineering schemes, such as quasi-resonant or resonant excitation, electrical carrier control, cavity enhancement, or Purcell engineering \cite{MudiPRL,Hauser2026,Karli2022,Albrechtsen2026}. While these approaches can substantially improve the optical properties of a QD, they also increase experimental complexity and can impose constraints on the applicability, scalability and network integration of the resulting devices.

An important outstanding challenge is therefore to realize telecom-band QDs that simultaneously provide high single-photon purity and strong photon indistinguishability under comparatively simple and practical excitation conditions, without relying on strong cavity enhancement. Such a platform would combine the intrinsic advantages of semiconductor quantum emitters with the low-loss transmission windows of optical fibers, while reducing the complexity associated with advanced excitation and photonic architectures. 

Several solid-state platforms have been investigated for the realization of single-photon sources operating in the telecom C-band, including Stranski–Krastanov (SK) grown In(Ga)As/InP QDs \cite{Benyoucef2013,Holewa2024,Phillips2024,Ge2024,Kaur2026}, In(Ga)As/GaAs QDs incorporating metamorphic buffer layers (MMBLs) \cite{Semenova2008,Paul2017,Wroski2021}, In(Ga)As/AlInGaAs/InP droplet QD~\cite{Ha2020,Karolina2026} and SK grown In(Ga)As/AlInGaAs/InP QDs \cite{Kim2025,Hauser2026,Behrends2026}. Among these approaches, the latter have demonstrated a Hong-Ou-Mandel (HOM) visibility of \(V_{3\,\mathrm{ns}}=91.7\%\) under longitudinal-acoustic (LA) phonon-assisted excitation. Under above-band excitation, however, the reported visibility reached only \((20.6 \pm 7)\%\), despite a Purcell enhancement factor of \(F_{\mathrm{P}}=(4.7 \pm 5)\). This contrast highlights the importance of the excitation scheme in determining photon indistinguishability~\cite{Yu2023}.

In this work, we demonstrate InAs/InP QD emitters operating in the telecom C-band that exhibit moderately high two-photon interference even under non-resonant, above-band excitation and without Purcell enhancement. By combining optimized metal-organic vapour-phase epitaxy (MOVPE) growth with deterministic in-situ electron-beam lithography (iEBL), we integrate individual QDs into circular Bragg grating (CBG) structures, enabling efficient photon extraction which facilitates characterizing their single-photon purity and indistinguishability. We quantify both multiphoton emission and the indistinguishability of consecutively emitted photons, obtaining a HOM visibility of \(V_{12\,\mathrm{ns}}= (35 \pm 2)\%\) for photons separated by the 12.5-ns excitation period. Placed in the context of reported telecom-band QD sources, these results establish a benchmark for above-band-excited InAs/InP QD-CBG emitters without cavity-induced Purcell enhancement. This work advances the development of practical solid-state quantum light sources for fiber-based quantum technologies by demonstrating telecom-band photon interference without requiring resonant excitation or Purcell enhancement.

\section{Methods}%

This work encompasses a comprehensive fabrication and characterization workflow, spanning numerical design and optimization, deterministic iEBL integration of individual QDs, nanofabrication, and detailed spectroscopic and quantum-optical characterization. In this section, we detail the numerical and experimental methodologies employed throughout the study, including FEM simulations and device design optimization, epitaxial growth, flip-chip integration, and nanostructure fabrication. We further describe the optical and structural characterization techniques, including microscopy, cathodoluminescence (CL), and micro-photoluminescence ($\mu$PL), as well as the quantum-optical measurements used to establish single-photon emission and quantify photon indistinguishability.

\subsection{Simulation and design optimization}

Numerical simulations were performed using the finite-element method (FEM) solver JCMsuite, which employs the FEM to solve Maxwell's equations~\cite{Pomplun2007}. Exploiting the rotational symmetry of the CBG, the structure was modeled using a two-dimensional axisymmetric geometry, substantially reducing the computational cost compared with full three-dimensional simulations. The computational mesh was chosen to adequately resolve the wavelength within each material, with local refinement applied near the dipole emitter and at material interfaces. Unless otherwise stated, the QD was positioned at the center of the CBG mesa. The refractive indices used in the simulations are summarized in Table~\ref{tab:table1}.

\begin{table}[b]
    \centering
    \caption{Refractive indices used in the simulations.}
    \label{tab:table1}
    \begin{tabular}{lcc}
        \hline
        Material & Refractive index & Reference \\
        \hline
        InP & 3.15 & \cite{Adachi1989} \\
        SiO$_2$ & 1.44 & \cite{Arosa:20} \\
        Al & $1.51 + 15.23\,\mathrm{i}$ & \cite{Rakic:98} \\
        \hline
    \end{tabular}
\end{table}

\begin{figure*}[!t]
    \centering
    \includegraphics[width=0.8\linewidth]{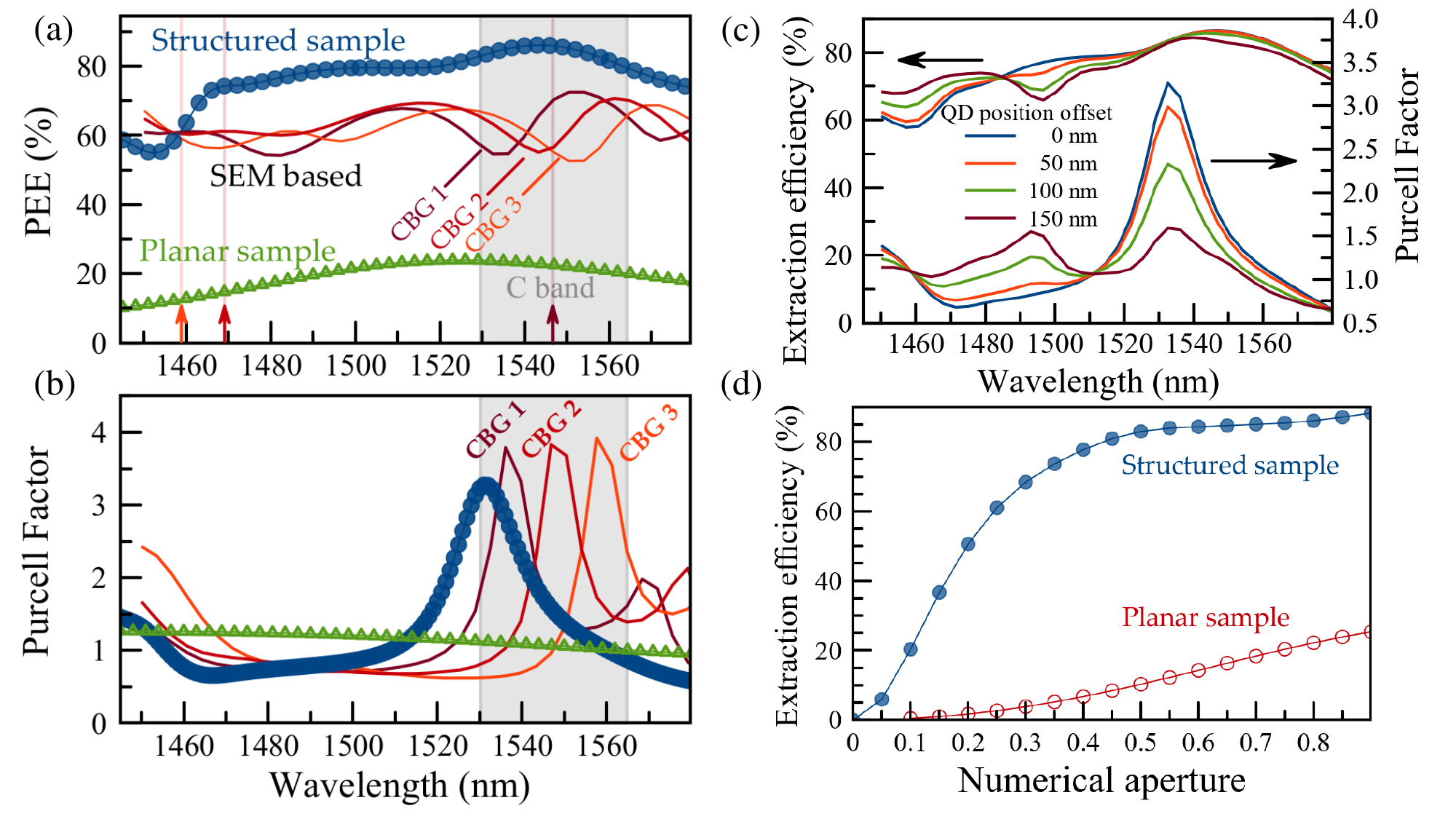}
    \caption{\textbf{Overview of the simulated PEE and $F_P$:} (a,b) Wavelength dependence of the PEE and $F_P$ for the planar as-grown reference (green triangles), the nominal simulated structure (blue circles), and the three fabricated devices CBG1, CBG2, and CBG3. Thin gray curves represent simulations of individual fabricated structures using SEM-measured dimensions, capturing fabrication-induced geometric variations; the black solid line denotes their average. Across the C-band (gray shading, 1530–1565 nm), the structured samples exhibit a $\sim$3–4$\times$ enhancement in both PEE and $F_P$ compared with the planar reference. The three representative devices (CBG1, CBG2, and CBG3; dark red, red, and orange, respectively) exhibit $F_P$ resonances within the C-band. The corresponding arrows in (a) indicate the wavelengths of the brightest measured spectral lines for each structure. (c) Simulated PEE (left axis) and $F_P$ (right axis) as a function of wavelength for increasing QD position offsets from the mesa center: \qty{0}{\nano\meter} (blue), \qty{50}{\nano\meter} (orange), \qty{100}{\nano\meter} (green), and \qty{150}{\nano\meter} (red). Increasing the lateral offset progressively reduced the $F_P$, whereas the PEE remains comparatively robust, highlighting the enhanced tolerance of photon extraction to QD-position uncertainty. (d) PEE as a function of collection numerical aperture (NA = 0–0.9) for the planar reference (red, open circles) and structured samples (blue, filled circles). The structured sample rapidly approaches a PEE of $\sim$83\% at NA = 0.5, while the planar reference exhibits a near-linear increase and reaches only $\sim$25\% at NA = 0.9.}
    \label{fig:simulation_Results} 
\end{figure*}

Numerical FEM simulations were performed over a wavelength range of \qtyrange{1430}{1580}{\nano\meter}; where no spectral dependence is shown, a wavelength of \qty{1550}{\nano\meter} was used. Design optimization was performed using the Bayesian optimization algorithm implemented in JCMsuite. The CBG geometry was optimized primarily to maximize photon extraction efficeincy (PEE), defined as the fraction of the total emitted power collected within the NA of the collection optics. The optimization target was a collection NA of 0.8 at a wavelength of \qty{1550}{\nano\meter}. The resulting structure consists of a central mesa with a radius of \qty{840}{\nano\meter}, surrounded by three concentric rings, each \qty{589}{\nano\meter} wide and separated by \qty{635}{\nano\meter}-wide trenches. The pattern is etched \qty{390}{\nano\meter} into the \qty{535}{\nano\meter}-thick InP layer, which is separated from a \qty{100}{\nano\meter} Al back reflector by a \qty{100}{\nano\meter} SiO$_2$ spacer. The QD layer is located \qty{243}{\nano\meter} below the top surface. The resulting structure and layer geometry are illustrated in Fig.~\ref{fig:SEM_image}(c) and Fig.~S1 of the Supporting Information (SI). 

For the optimized geometry, the simulated PEE exceeds 80\% throughout most of the C-band, reaching approximately 83\% near the center of the band, compared with approximately 20--25\% for the planar reference. This corresponds to an enhancement of approximately $3$--$4\times$ across the C-band and demonstrates the efficient redistribution of the QD emission into the collection cone of the CBG. The corresponding $F_P$ remains comparatively modest, reaching a maximum slightly above 3 around \qty{1530}{\nano\meter} with a spectral width of approximately \qty{18}{\nano\meter}. The limited Purcell enhancement is a direct consequence of optimizing the structure primarily for photon extraction rather than for strong modification of the local density of optical states. In particular, maximizing PEE and Purcell enhancement simultaneously generally involves a trade-off, and the three-ring geometry considered here was selected to prioritize broadband and efficient photon collection over large Purcell enhancement.

The wavelength-dependent performance of the nominal design and the fabricated CBGs is shown in Fig.~\ref{fig:simulation_Results}(a,b). The fabricated devices CBG1, CBG2, and CBG3 exhibit resonances within the C-band, while simulations based on their experimentally determined SEM dimensions account for fabrication-induced geometric variations. The good agreement between the nominal and fabricated structures demonstrates that the optimized geometry retains its high extraction performance despite geometrical deviations introduced during fabrication. The arrows in Fig.~\ref{fig:simulation_Results}(a) indicate the wavelengths of the brightest measured emission lines used for the subsequent optical characterization. To assess the robustness of the design against the finite positioning accuracy of iEBL integration, we further investigated the effect of lateral displacement of the QD from the mesa center [Fig.~\ref{fig:simulation_Results}(c)]. Although iEBL enables deterministic positioning, residual sample drift during the fabrication process can result in lateral offsets of several tens of nanometers \cite{Donges2024}. This effect is particularly pronounced for telecom-band QDs, where the relatively low signal-to-noise ratio of InGaAs linear-array detectors necessitates longer integration times, thereby increasing susceptibility to sample drift and compromising the lateral positioning accuracy of the QD \cite{Limame2026}.

\begin{figure*}[!t]
    \centering
    \includegraphics[width=1\linewidth]{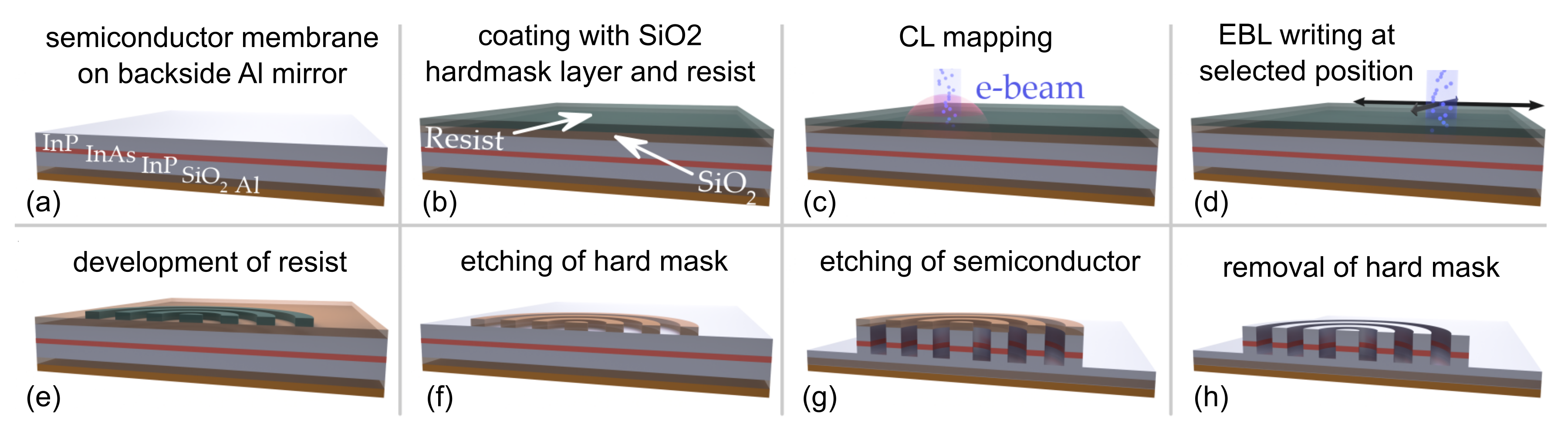}
    \caption{\textbf{Schematic illustration of the sample structure and fabrication workflow:} (a) Initial flip-chip-bonded sample comprising the InAs/InP QD active region, an SiO$_2$ spacer layer, and an Al back reflector. (b) Deposition of an SiO$_2$ hard mask by PECVD followed by spin-coating of CSAR resist to prepare the sample for selective nanostructuring. (c) Low-dose CL mapping to identify individual QDs emitting within the telecom C-band. (d,e) Alignment of the selected QD to the center of the CBG, followed by iEBL and resist development. (f,g) Transfer of the CBG pattern into the SiO$_2$ hard mask by RIE, followed by ICP-RIE to etch the underlying InP layer containing the InAs QDs. (h) Post-etch cleaning to remove the SiO$_2$ hard mask and residual resist, yielding the final CBG structure precisely aligned to the selected QD.}
    \label{fig:sample_processing_steps}
\end{figure*}

The simulations show that the PEE remains remarkably robust against such displacement: even for offsets up to \qty{150}{\nano\meter}, the extraction efficiency remains comparatively high. In contrast, the Purcell enhancement is considerably more sensitive to the QD position, with increasing displacement progressively suppressing the resonance. This difference reflects the distinct physical requirements of efficient far-field emission and strong local field enhancement. While the extraction efficiency is therefore tolerant to realistic positioning errors, QD displacement can still modify the far-field emission pattern and polarization, which may become relevant for efficient coupling to single-mode photonic channels. Finally, we evaluated the dependence of the PEE on the collection NA at \qty{1550}{\nano\meter} [Fig.~\ref{fig:simulation_Results}(d)]. The CBG rapidly redirects the emission into the collection cone, reaching approximately 60\% at an NA of 0.25 and approaching a plateau of approximately 83\% for NA = 0.5. In contrast, the planar reference exhibits a substantially weaker and more gradual increase, reaching only approximately 25\% at NA = 0.9. Thus, the optimized CBG provides efficient photon extraction without requiring extremely high-NA collection optics, demonstrating its suitability for practical and scalable single-photon collection architectures.

\subsection{Epitaxial growth and nanofabrication}

The devices investigated in this work are based on near critical Stranski-Krastanov self-assembled InAs QDs embedded in an InP matrix\cite{Berdnikov2024NearCriticalSK}. The QDs were grown between two InP layers with thicknesses of \qty{243}{\nano\meter} and \qty{292}{\nano\meter} on an InP (100) substrate. To enable the subsequent flip-chip integration and formation of a bottom reflector, the upper InP layer was capped by a \qty{100}{\nano\meter} SiO$_2$ spacer and a \qty{100}{\nano\meter} Al mirror. The processed structure was subsequently flip-chip bonded to a Si carrier, resulting in a total InP thickness of \qty{535}{\nano\meter} in the final device. The fabrication process is described in detail in \cite{Holewa2022}, and the resulting layer configuration is schematically illustrated in Fig.~\ref{fig:sample_processing_steps}(a).

The CBGs were fabricated using our deterministic iEBL process that combines low-temperature CL mapping, QD localization, and nanostructure patterning within a single low-temperature iEBL session~\cite{Rodt2021}. This approach enables the position of an individual QD to be directly correlated with the subsequently written CBG, thereby avoiding the need for conventional marker-based alignment. The complete fabrication sequence is summarized in Fig.~\ref{fig:sample_processing_steps}. 

Because the CSAR~62 (ARP~6200.18) electron-beam resist does not provide sufficient selectivity for direct etching of InP, an additional SiO$_2$ hard mask was employed. First, a \qty{200}{\nano\meter} SiO$_2$ layer was deposited by plasma-enhanced chemical vapor deposition (PECVD), followed by the application of a thin hexamethyldisilazane (HMDS) adhesion layer. CSAR~62 resist was then spin-coated at 4000~rpm, yielding a thickness of approximately \qty{780}{\nano\meter} [Fig.~\ref{fig:sample_processing_steps}(b)]. The chosen layer thicknesses provide the required etch resistivity for transferring the CBG pattern first into the SiO$_2$ hard mask and subsequently into the underlying InP layer.

\begin{figure*}
    \centering
    \includegraphics[width=1\linewidth]{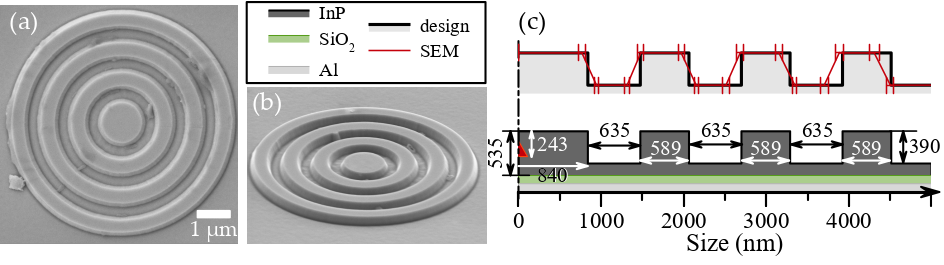}
    \caption{\textbf{Characterization of the fabricated CBGs:} (a) Top-view SEM image of a fabricated CBG. The ring edges exhibit pronounced contrast, with brighter segments revealing the formation of tilted sidewalls during the fabrication process. (b) Cross-sectional SEM image confirming the resulting non-vertical ring profiles and characteristic sidewall tilt. (c) Comparison between the optimized and experimentally realized geometries. The ideal design is shown by the black lines, while the red curves represent the measured profiles extracted from SEM images. The black line corresponds to the targeted vertical sidewalls, whereas the red curve denotes the average sidewall profile obtained across the fabricated structures. Red vertical error bars indicate the corresponding structure-to-structure variation in the measured sidewall slope.}
    \label{fig:SEM_image}
\end{figure*}

The deterministic QD localization was performed at cryogenic temperature using the integrated CL--EBL system. A focused electron beam was raster-scanned across the sample, and the resulting QD luminescence was spectrally resolved in the near-infrared range. This enables spatial mapping of individual emitters and, importantly, identification of QDs emitting within the target telecom C-band prior to nanostructure fabrication [Fig.~\ref{fig:sample_processing_steps}(c)].

After identifying the target emitters, the CBG patterns were aligned to the extracted QD positions and directly written into the CSAR resist using the same iEBL system [Fig.~\ref{fig:sample_processing_steps}(d,e)]. The use of iEBL minimizes the time between QD localization and patterning, thereby reducing errors associated with sample drift at cryogenic temperature. This is particularly important for deterministic positioning, where even small lateral displacements between QD identification and nanostructure fabrication can affect the optical performance of the resulting device. The CSAR~62 resist exhibits a non-monotonic dose response, with a positive tone at intermediate doses and a transition to negative tone at higher doses. For the present fabrication process, a CL mapping dose of \qty[per-mode=symbol]{4}{\milli\coulomb\per\centi\meter\squared} was used to maintain the resist in the positive-tone regime, whereas a patterning dose of \qty[per-mode=symbol]{32}{\milli\coulomb\per\centi\meter\squared} was used to define the CBG structures in the negative-tone regime. Further details on the dose-dependent response of CSAR~62 can be found in Ref.~\cite{Kaganskiy_2016}.

After electron-beam exposure, the sample was removed from the iEBL system and developed in AR600-546. The resulting resist pattern was subsequently transferred into the SiO$_2$ hard mask by reactive-ion etching (RIE) using SF$_6$ and Ar [Fig.~\ref{fig:sample_processing_steps}(f)]. The etching was performed at gas flow rates of 16~sccm SF$_6$ and 2~sccm Ar, with an RF power of \qty{50}{\watt} and a chamber pressure of \qty{3}{\pascal}. The remaining CSAR resist was then removed by immersing the sample in dimethyl sulfoxide (DMSO) overnight.

The patterned SiO$_2$ layer subsequently served as an etch mask for transferring the CBG geometry into the underlying InP [Fig.~\ref{fig:sample_processing_steps}(g)]. The InP was etched to a target depth of \qty{390}{\nano\meter} using an inductively coupled plasma reactive-ion etching (ICP-RIE) process with equal flow rates of 50~sccm Cl$_2$ and Ar. The RF power was set to \qty{100}{\watt} at a chamber pressure of \qty{0.5}{\pascal}. Following the InP etch, the residual SiO$_2$ hard mask was removed by wet etching in buffered oxide etchant (BOE), yielding the final CBG structure [Fig.~\ref{fig:sample_processing_steps}(h)].

The resulting nanostructure geometry was characterized by scanning electron microscopy (SEM). Representative top- and cross-sectional SEM images are shown in Fig.~\ref{fig:SEM_image}(a,b). The ring edges exhibit pronounced contrast variations, with locally enhanced brightness associated with the non-vertical sidewalls produced during the etching process. The cross-sectional image directly confirms the resulting tapered profile. To quantify the deviation from the nominal design, the experimentally obtained sidewall profiles were extracted from SEM images and compared with the optimized geometry [Fig.~\ref{fig:SEM_image}(c)]. The idealized design is represented by the black curves, while the red curves correspond to the experimentally measured profiles. The measured structures exhibit a systematic sidewall tilt relative to the nominal vertical profile, together with a finite structure-to-structure variation, as indicated by the corresponding red error bars in Fig.~\ref{fig:SEM_image}(c). These experimentally determined geometrical deviations were subsequently incorporated into the numerical simulations to account for fabrication-induced variations in the optical response, as shown in Fig. \ref{fig:simulation_Results}.

Together, the iEBL process and subsequent nanofabrication enable the realization of CBG structures with controlled positioning relative to individual pre-selected telecom-band QDs. The combination of CL-based emitter localization and SEM-based structural characterization further provides the experimental basis for correlating the optical response of each device with both the actual QD position and the realized CBG geometry.

\section{Spectroscopic and Quantum-Optical Characterization}

Optical characterization of the QDs was performed at cryogenic temperatures (T \(\sim 4~\mathrm{K}\)) using $\mu$PL spectroscopy and time-correlated single-photon detection. The setup enabled measurements of PEE, radiative dynamics, single-photon purity, and photon indistinguishability through time-resolved PL (TRPL), Hanbury Brown and Twiss (HBT), and HOM experiments. Further details of the optical setup, excitation conditions, detection system, and measurement configurations are provided in Section S1 of the SI.

\subsection{Emission spectra and photon extraction efficiency}

\begin{figure}
    \centering
    \includegraphics[width=1\linewidth]{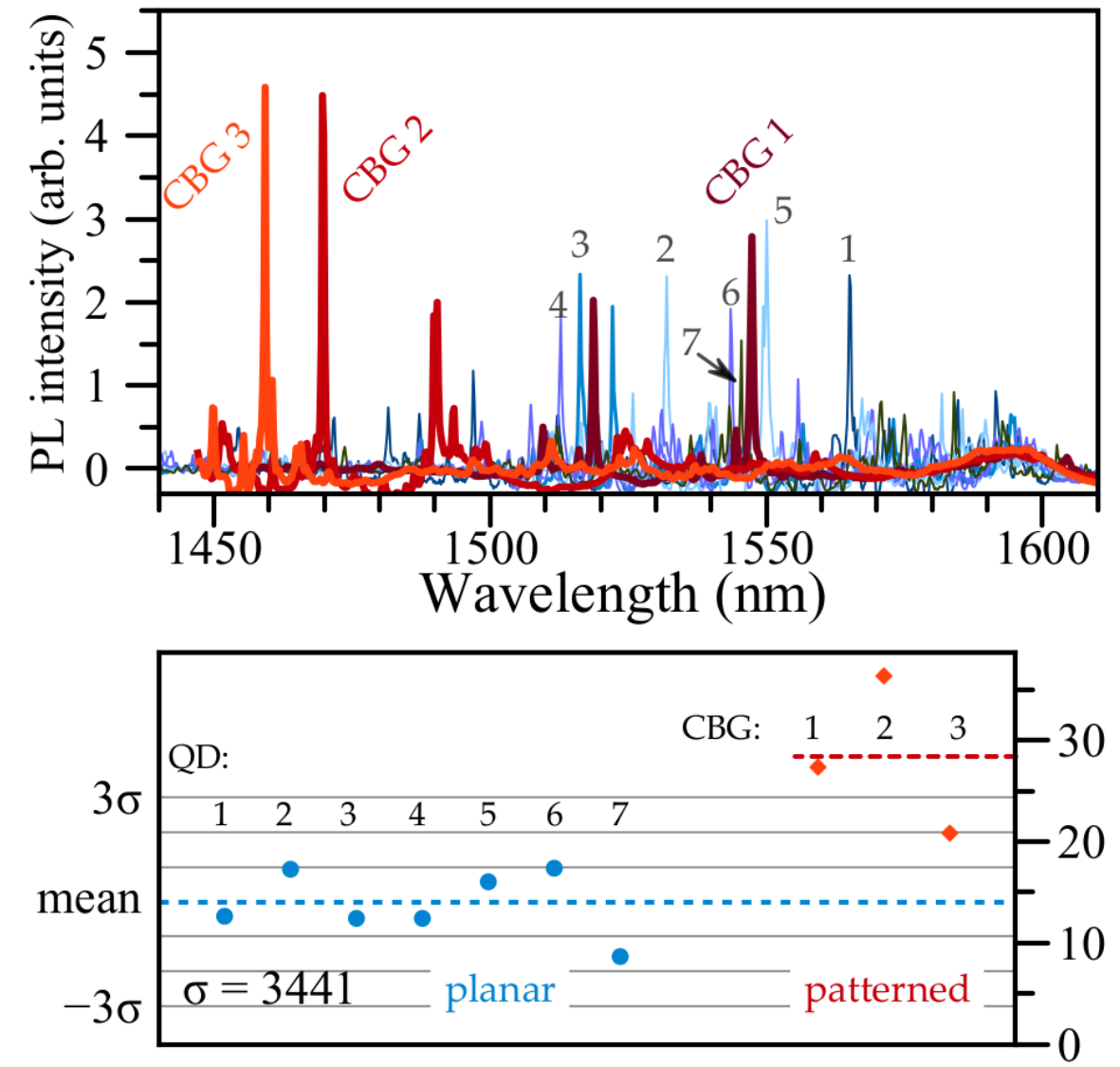}
    \caption{\textbf{PEE enhancement of QDs embedded in CBG structures:} (a) Background-corrected $\mu$PL spectra for seven planar reference QDs (1–7, blue) and three QD-CBG devices (CBG1, CBG2, and CBG3, red). (b) Statistical comparison of the integrated $\mu$PL intensities for the planar reference QDs and the three CBG structures. The CBG-embedded QDs (red diamonds) exhibit emission intensities exceeding the $+3\sigma$ threshold of the planar reference distribution (blue circles), corresponding to an approximately $2\times$ enhancement in brightness.}
    \label{fig:PL_spectra}
\end{figure}

To estimate the photon extraction enhancement by the CBGs, we compare the $\mu$PL spectra of QDs embedded in the planar reference sample with those of QDs integrated into the CBG structures recorded under CW laser excitation at 785~nm at saturation. Representative spectra from both configurations are shown in Fig.~\ref{fig:PL_spectra}(a). For the planar reference, spectra were acquired from seven randomly selected QD positions, while the CBG measurements correspond to the QDs integrated into devices CBG1, CBG2, and CBG3. The comparison provides a direct assessment of the enhancement in collected emission enabled by the CBG structures relative to the unpatterned reference.

The QD–CBG devices are represented by thick reddish curves, whereas the seven randomly selected QDs in the planar reference region are shown as thin bluish curves, with each spectrum individually labeled. Notably, the emission peaks of devices CBG2 and CBG3 exhibit substantially higher amplitudes than those of all measured planar QDs, providing a clear indication of enhanced PEE by the CBG structures. For quantitative analysis, the emission lines were fitted using Lorentzian line shapes. The QD–CBG emitters exhibit a mean full width at half maximum (FWHM) of $\qty{0.45}{\nano\meter}$ with a standard deviation of $\qty{0.13}{\nano\meter}$, compared with $\qty{0.53}{\nano\meter}$ and $\qty{0.16}{\nano\meter}$, respectively, for the planar QDs. The comparable linewidths indicate that the CBG integration does not introduce a substantial spectral broadening of the emitters.

The brightness of each QD was quantified as the integrated area under the fitted emission peak. The resulting brightness values are presented in Fig.~\ref{fig:PL_spectra}(b) after normalization to the planar reference population using a standard-score transformation. For the seven planar QDs, the mean brightness is \qty{14067}{\arbunit} with a standard deviation of \qty{3441}{\arbunit}. Each measured brightness $B$ was therefore expressed as

$$
z = \frac{B-\langle B_{\mathrm{planar}}\rangle}{\sigma_{\mathrm{planar}}},
$$

yielding a dimensionless measure of the deviation from the planar reference population. The blue circular markers represent the individual planar QDs, while the red diamond markers correspond to the QD--CBG devices. The blue dashed line denotes the mean brightness of the planar QDs, and the gray dashed lines indicate $\pm1$ standard deviation from this mean.

The integrated brightness values obtained for devices CBG1, CBG2, and CBG3 are \qty{27362}{\arbunit}, \qty{36373}{\arbunit}, and \qty{20847}{\arbunit}, respectively. All three QD--CBG devices exhibit brightness values above the planar reference distribution, demonstrating a significant enhancement in the collected emission. The mean brightness of the three CBG-integrated emitters, indicated by the red dashed line in Fig.~\ref{fig:PL_spectra}(b), is approximately twice that of the planar reference QDs. This systematic enhancement provides direct experimental evidence of the increased photon collection enabled by the CBG structures.

The $\mu$PL comparison demonstrates a pronounced enhancement in the collected emission from QDs integrated into CBG structures relative to those in the planar reference sample. Across the investigated devices, the average integrated brightness is approximately doubled upon CBG integration, indicating a substantial improvement in photon collection efficiency, where the deviation from the theoretically predicted enhancement is explained by a slightly off-center position of the QDs in the CBG. In addition to enhancing the desired QD emission, the CBG structures substantially suppress the broadband background emission, thereby improving the signal-to-background ratio and providing favorable conditions for high-quality single-photon characterization.

Beyond the spectral comparison, we performed a comprehensive set of complementary measurements to further characterize the emission properties of the QD--CBG devices, including TRPL, polarization-resolved PL, and temperature-dependent PL. These measurements provide additional insight into the fundamental optical properties of the emitters and establish the characteristics relevant to their operation as single-photon sources. Although these properties are not the primary focus of the present study, they are relevant for a complete characterization of the devices. A detailed description of the corresponding measurements and results is provided in the SI.

\begin{figure*}
    \centering
    \includegraphics[width=1\linewidth]{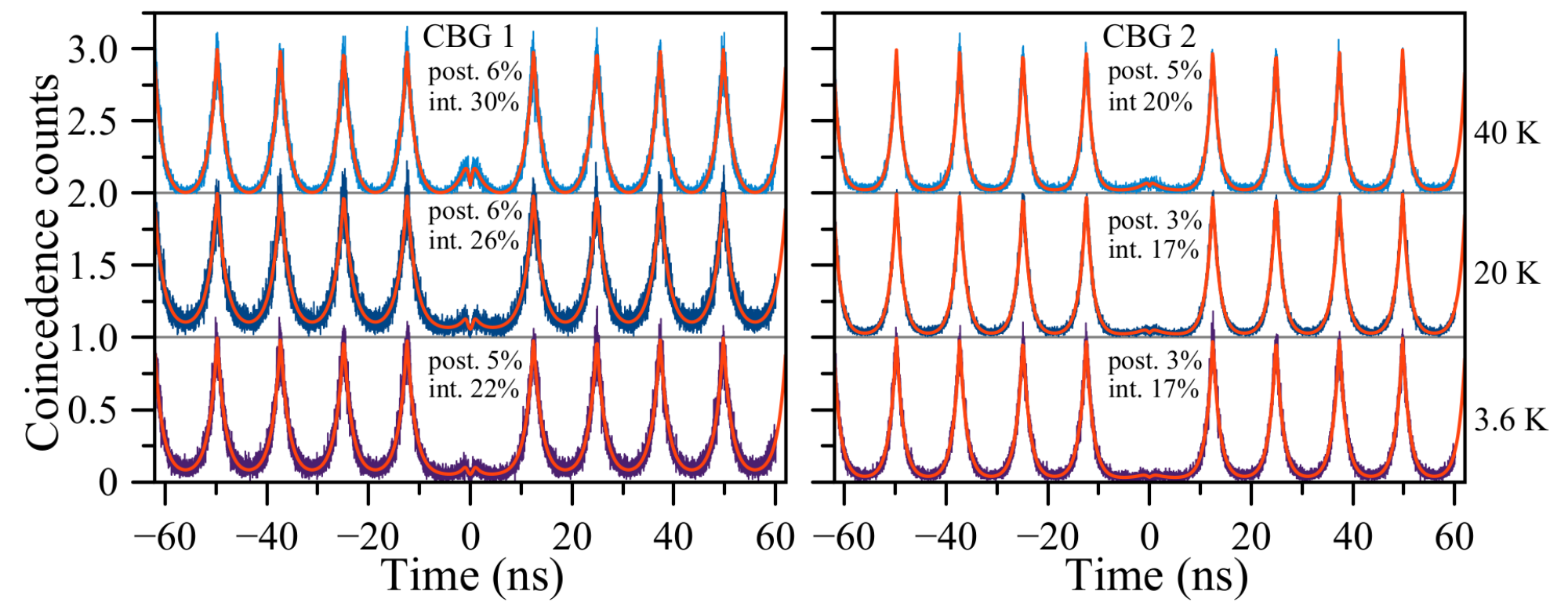}
    \caption{\textbf{Second-order auto-correlation measurements of CBG-embedded QDs under pulsed non-resonant excitation:} HBT measurements were performed at $\lambda=800$ nm for varying temperatures. The measurements for devices CBG1 and CBG2 are shown in the left and right columns, respectively. For each structure, $g^{(2)}(\tau)$ is presented at 3.6 K (bottom), 20 K (middle), and 40 K (top). Experimental coincidence histograms (blue) are fitted with the corresponding theoretical model (red). The pronounced suppression of coincidences at zero time delay ($\tau=0$ ns) demonstrates strong photon antibunching and confirms the single-photon character of the emission across the investigated temperature range. The corresponding post-selected (proc) and integrated (int) $g^{(2)}(0)$ values are indicated for each measurement.}
    \label{fig:g2_spektra}
\end{figure*}

Power- and polarization-dependent measurements were performed on devices CBG1 and CBG2, which are considered exclusively in the following analysis to determine the nature of the observed emission lines. The brightest emission lines, centered at \qty{1544}{\nano\meter} for CBG1 and \qty{1470}{\nano\meter} for CBG2, both exhibit an approximately linear dependence on excitation power ($P^1$), characteristic of single-exciton emission. Furthermore, polarization-resolved measurements reveal FSS values of $(52 \pm 2)~\mu eV$ and $(74 \pm 5)~\mu eV$ for CBG1 and CBG2, respectively. The combination of near-linear power dependence and finite FSS provides strong evidence that the observed emission lines originate from neutral excitonic transitions.

\subsection{Temperature dependent performance}

To assess the suitability of the QD-CBG devices for operation at elevated temperatures, which are advantageous from the perspective of practical implementation, scalability, and thermal management, we performed temperature-dependent optical and quantum-optical measurements on the two devices CBG1, CBG2 and CBG3. Measurements were conducted over a temperature range of approximately \qtyrange{4}{55}{\kelvin}, extending beyond the operating range of compact Stirling cryocoolers and thereby providing insight into the thermal stability of the devices under technologically relevant conditions\cite{Schlehahn2015,Aigner2026}. Representative $\mu$PL spectra acquired at different temperatures are presented in Section~S\textit{3.2} of the SI.

Temperature-dependent PL reveals thermally activated carrier redistribution and escape in the CBG-integrated QDs, manifested as emission redshift, linewidth broadening, and intensity quenching (Figures S4-5). Analysis of the dominant excitonic transitions in CBG2 and CBG3 identifies two characteristic activation temperatures, varying between emitters and excitation conditions. CBG1 exhibits a distinct temperature dependence, characterized by spectral broadening and a comparatively weak reduction in the integrated emission intensity. The broadening may reflect increased exciton–phonon interactions, while the temperature-dependent intensity quenching is consistent with thermally activated carrier escape from the QD into non-radiative states. The extracted activation temperatures therefore characterize the temperature dependence of these thermally activated processes rather than carrier redistribution between neighboring emission states. Further details of the temperature-dependent spectra and extracted parameters are provided in Section~S\textit{3.2} of the SI.

A key figure of merit for a single-photon source is the suppression of multiphoton emission, which can be quantified through the second-order autocorrelation function, $g^{(2)}(\tau)$, measured using a HBT configuration. Figure~\ref{fig:g2_spektra} presents the experimental $g^{(2)}(\tau)$ for emitters CBG1 and CBG2 under non-resonant pulsed excitation at 800~nm with a repetition rate of 80~MHz at saturation. Measurements were performed at 3.6, 20, and 40~K. The correlation histograms exhibit a pronounced suppression of the zero-delay peak for both CBG-QD devices at all temperatures, demonstrating strong antibunching and confirming their single-photon character.

The strongest suppression is observed for CBG2 at 3.6~K. The zero-delay correlation extracted from a fit to the correlation function is $g^{(2)}(0)= (0.03 \pm 0.01)$. We refer to this quantity as the \textit{post-selected} value. The fit accounts for the temporal structure of the pulsed correlation signal and provides a measure of the multiphoton contribution localized around zero delay. For comparison, we also quantify the central-peak contribution directly from the measured histogram. Specifically, the area of the zero-delay peak within a 12-ns integration window is normalized to the average area of the adjacent side peaks. This procedure yields an \textit{integrated} value of $g^{(2)}(0) = (0.17 \pm 0.03)$ for CBG2 at 3.6~K. 

The difference between the post-selected and integrated values is physically significant. The integrated metric is sensitive not only to true multiphoton emission within a single excitation cycle, but also to temporal broadening and correlations extending over a substantial fraction of the 12.5-ns excitation period. In particular, when the emitter lifetime is comparable to the temporal separation between subsequently excited states, the correlation peaks broaden and partially overlap. This effect increases the integrated central-peak area even when the short-time antibunching remains strong. Consequently, the post-selected $g^{(2)}(0)$ provides a more localized measure of the multiphoton probability, whereas the integrated metric captures additional temporal correlations present in the experimentally measured histogram. 

For CBG1, the corresponding values are $g^{(2)}(0) = (0.05 \pm 0.02)$ from the fit and $g^{(2)}(0) = (0.22 \pm 0.05)$ from the integrated analysis at 3.6~K. We attribute this behavior primarily to carrier recapture following non-resonant excitation. Under non-resonant excitation, a single laser pulse generates a population of carriers in the surrounding semiconductor. After the QD emits a photon, residual carriers may remain available for subsequent capture by the QD. A second electron--hole pair can therefore be captured within the same excitation cycle, giving rise to a second radiative recombination event. This mechanism is particularly important because the laser repetition period is $12.5$~ns, whereas carrier capture and relaxation occur on much shorter timescales. As a result, the QD can emit more than one photon following a single excitation pulse. In the correlation histogram, genuine antibunching is manifested primarily through the strong suppression at $\tau=0$, while carrier recapture can produce additional correlations at finite delays within the same excitation period. The latter contribution increases the integrated central-peak area and therefore leads to a larger integrated value than the $g^{(2)}(0)$ obtained from the local fit.

\begin{figure*}[t]
    \centering
    \includegraphics[width=1\linewidth]{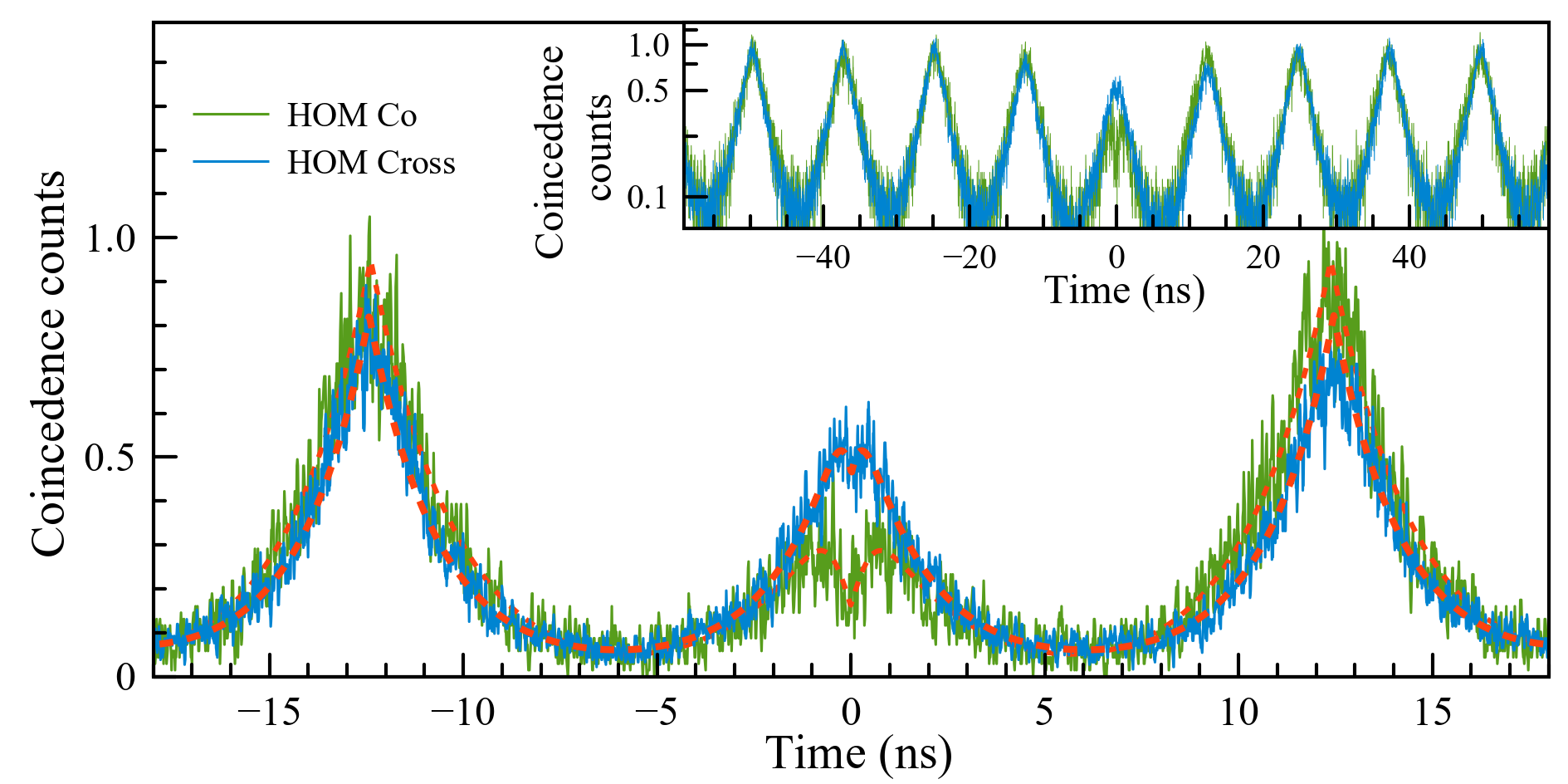}
    \caption{\textbf{Two-photon interference of CBG-embedded QD CBG1:} The main panel shows the normalized coincidence counts as a function of temporal delay for structure CBG1. A pronounced suppression of coincidences at zero delay ($\tau=0$ ns) is observed for co-polarized photons (green, HOM Co), while a weaker suppression remains visible for cross-polarized photons (blue, HOM Cross). The corresponding measurements are fitted by the red dotted curves. The inset provides a wider view of the periodic two-photon interference pattern. From the fitted coincidence profiles, a HOM visibility of \qty[separate-uncertainty = true]{35 \pm 2}{\percent} is extracted, demonstrating clear two-photon interference from the CBG-embedded QD.}
    \label{fig:HOM_spectra}
\end{figure*}

The temperature dependent optical studies further support this interpretation. At $20$~K, CBG2 retains a similar degree of antibunching, whereas CBG1 exhibits a modest change, with post-selected and integrated values of approximately $0.06$ and $0.26$, respectively. Upon increasing the temperature to $40$~K, the single-photon performance deteriorates more noticeably. For CBG2, the post-selected value increases to approximately $0.05$, with an integrated value of about $0.20$. For CBG1, the corresponding values are approximately $0.06$ and $0.30$. The temperature dependence can be understood in terms of thermally activated carrier dynamics. Increasing the temperature can enhance carrier escape and redistribution within the surrounding semiconductor environment, modify the carrier-capture dynamics, and increase the probability of re-excitation or recapture processes. These effects provide additional pathways for sequential photon emission following a single excitation pulse, thereby reducing the observed antibunching. The stronger temperature sensitivity of CBG1 compared with CBG2 suggests that the two emitters experience different local carrier environments or confinement conditions, which modify their susceptibility to thermally activated recapture.

Overall, both QD-CBGs maintain pronounced antibunching over the investigated temperature range, with CBG2 exhibiting the lowest multiphoton contribution and the most robust single-photon performance.

\subsection{Photon indistinguishability}

Photon indistinguishability is essential for solid-state quantum technologies, enabling interference-based quantum information processing and networking. We quantify the indistinguishability of consecutively emitted photons through HOM interference. Co-polarized measurements allow two-photon interference, while cross-polarized measurements serve as a distinguishable reference to isolate quantum interference from classical temporal overlap.

Figure~\ref{fig:HOM_spectra} shows the HOM correlation histograms measured for the neutral exciton of the CBG1 device, emitting near the center of the telecom C-band. The cross-polarized measurement, shown in blue, exhibits the characteristic series of correlation peaks corresponding to photons generated by consecutive excitation cycles. Since the two photons are rendered distinguishable by their orthogonal polarization states, no HOM suppression is expected at zero delay. The co-polarized measurement, shown in green, provides the corresponding indistinguishable-photon configuration. A pronounced reduction of the central coincidence peak is observed for the co-polarized configuration compared with the cross-polarized reference. This suppression at $\tau=0$ constitutes direct evidence of two-photon interference and demonstrates that photons emitted by the QD during successive excitation cycles retain a substantial degree of indistinguishability. The observation is particularly significant for a non-resonantly excited QD, where excitation-induced carrier relaxation and fluctuations in the local charge environment can introduce temporal and spectral variations between consecutively emitted photons.

\begin{table*}[t]
\centering
\caption{Comparison of quantum-dot single-photon sources.
$F_{\textrm{P}}$ -- Purcell factor; Exc. state -- excitonic state.}
\label{tab:comparison}
\footnotesize
\setlength{\tabcolsep}{3pt}
\renewcommand{\arraystretch}{1.2}

\begin{tabularx}{\textwidth}{
>{\raggedright\arraybackslash}p{2.0cm}
>{\centering\arraybackslash}p{1.4cm}
>{\centering\arraybackslash}p{1.3cm}
>{\centering\arraybackslash}p{1.2cm}
>{\centering\arraybackslash}p{1.2cm}
>{\centering\arraybackslash}p{1.2cm}
>{\centering\arraybackslash}p{2.5cm}
>{\centering\arraybackslash}p{1.8cm}
>{\centering\arraybackslash}p{1.5cm}
>{\centering\arraybackslash}p{1.5cm}
}
\hline
QD Type & Growth & Device & Excitation & $F_{\textrm{P}}$ &
Exc. state & $V_{\mathrm{HOM}}$ & $T_1$ &
$\Delta E_{\mathrm{FWHM}}$ & Refs. \\
\hline

InAs/InP & MOVPE & CBG/Al & Above band & 1 &
$X^{+/-}$ & $V_{12\,\mathrm{ns}}=(35\pm2)\%$ &
$(1400\pm50)$ ps & $41.0\pm4.0$ $\mu$eV & This work \\

InAs/MMB & MOVPE & CBG/Au & LA & $4.3\pm0.8$ &
$X^+$ & $V_{\mathrm{fit}}=34.9\%$ &
500 ps & 20.0 $\mu$eV & \cite{Joos2024} \\

InAs/InP & MOVPE & Mesa/Al & TPE & 1 &
$XX$ & $V_{4\,\mathrm{ns}}=35\%$ &
340 ps & 47 $\mu$eV & \cite{Vajner2024} \\

InAs/InP & MOVPE & CBG/Al & LO & 5 &
$X^{+/-}$ & $V_{\mathrm{fit}}=19.3\%$ &
530 ps & 87 $\mu$eV & \cite{Holewa2024} \\

InAs/InAlGaAs & MBE & CBG/Au & Above band &
$4.7\pm0.5$ & $X^{+/-}$ &
$V_{3\,\mathrm{ns}}=(20.6\pm7)\%$ &
257.5 ps & 28.9 $\mu$eV & \cite{Hauser2026} \\

InAs/InAlGaAs & MBE & CBG/Au & LA &
$4.7\pm0.5$ & $X^{+/-}$ &
$V_{3\,\mathrm{ns}}=91.7\%$ &
257.5 ps & 28.9 $\mu$eV & \cite{Hauser2026} \\

\hline
\end{tabularx}
\end{table*}

The HOM visibility is obtained by comparing the integrated coincidence probability of the central peak in the co- and cross-polarized configurations. To provide a more localized measure of the interference around zero delay, we  integrate the coincidence histograms over the interval $-6~\mathrm{ns}\leq\tau\leq6~\mathrm{ns}$. This analysis yields a HOM visibility of \[ V_{\mathrm{HOM}} = (35 \pm 2)\%. \] 

The influence of the temporal integration window on the extracted visibility is investigated in the SI (Fig. S6). As the integration window is progressively extended, the visibility decreases from $(70 \pm 18)\%$ for the shortest window to $(35 \pm 2)\%$ when integrating over the full laser repetition period of $\qty{12.5}{\ns}$. This reduction reflects the increasing contribution of coincidence events outside the central interference region to the integrated coincidence probabilities. The resulting visibility therefore depends on the selected integration window, which is an important consideration when comparing measured HOM visibilities.

The observed interference demonstrates a substantial degree of mutual coherence between photons emitted from different excitation cycles. In an idealized description, the HOM visibility is directly related to the overlap of the two single-photon wave packets. Any spectral wandering, temporal jitter, phonon-induced dephasing, or fluctuations of the QD environment reduce this overlap and therefore suppress the interference contrast.

The observation of $V_{\mathrm{HOM}}=(35 \pm 2)\%$ under non-resonant excitation and without Purcell enhancement is particularly notable. Non-resonant excitation generally introduces additional carrier relaxation pathways and increases the sensitivity of the emitter to charge and environmental fluctuations, which can degrade photon indistinguishability. The substantial HOM interference observed here therefore indicates that the CBG1 device maintains a comparatively stable optical transition and coherent emission despite the non-resonant excitation scheme.

\section{Comparison of quantum-dot single-photon sources}

To benchmark our results, we compare the measured photon indistinguishability with recently reported QD single-photon sources, as summarized in Table~\ref{tab:comparison}. Our measured two-photon interference visibility of $V_{12,\mathrm{ns}}=(35 \pm 2)\%$ is comparable to values reported for several state-of-the-art QD sources, despite the relatively simple experimental conditions used in this work. In particular, our source is operated under above-band excitation and does not rely on a pronounced Purcell enhancement, with a $F_P$ close to unity. This is in contrast to several of the compared sources, where resonant or quasi-resonant excitation schemes and/or significant Purcell enhancement are employed to improve the photon coherence.

The comparison is particularly relevant when considering the material platform and growth method. Our InAs/InP QDs are grown by MOVPE, whereas a number of the highest reported indistinguishability values in Table~\ref{tab:comparison} are obtained from structures based on MBE, often in combination with Purcell-enhancing cavities and more selective excitation schemes. MOVPE offers a technologically established route to scalable epitaxial growth of InP-based semiconductor structures and is well suited to integration with mature InP photonic platforms for telecom-wavelength applications \cite{Wang2024}, but the resulting QD properties can differ from those obtained using MBE, including variations in charge noise, structural uniformity, and optical linewidth \cite{Smoka2021}. These material and growth-related differences should therefore be considered when comparing the absolute values of $V_{\mathrm{HOM}}$.

The measured linewidth of $\Delta E_{\mathrm{FWHM}}= (41.0 \pm 4.0)~\mu\mathrm{eV}$ and the radiative lifetime of $T_1= (1400 \pm 50)~\mathrm{ps}$ indicate that there is substantial room for further improvement. The absence of significant Purcell enhancement, the relatively long radiative lifetime of semiconductor QDs increases their susceptibility to pure dephasing and environmental spectral fluctuations, which can reduce the indistinguishability of emitted photons when these processes occur on timescales comparable to the radiative lifetime. A reduction of the radiative lifetime through cavity enhancement would therefore be expected to increase the coherence of the emitted photons \cite{Rickert2024}. Similarly, moving from above-band excitation towards resonant or quasi-resonant excitation could suppress charge fluctuations and excess carriers generated in the surrounding semiconductor, thereby reducing spectral diffusion and improving the indistinguishability \cite{Hauser2026}.

Importantly, the present result demonstrates that a measurable degree of photon indistinguishability can already be obtained from MOVPE-grown InAs/InP QDs using above-band excitation and without significant Purcell enhancement. This provides a useful baseline for subsequent improvements. Combining the present platform with improved charge stabilization, more selective excitation, and optimized photonic structures could provide a direct route towards substantially higher indistinguishability while retaining the scalability advantages of the MOVPE/InP material system.


\section{Conclusions}\label{sec5}

In conclusion, we demonstrate deterministic iEBL integration of MOVPE-grown InAs/InP QDs into CBG resonators for telecom C-band single-photon generation. Under non-resonant above-band excitation and without Purcell enhancement, the integrated emitters exhibit a post-selected single-photon purity of (\(0.95 \pm 0.02\)) and a raw Hong–Ou–Mandel interference visibility of \((35 \pm 2)\%\) for photons separated by 12.5 ns and up to $(70 \pm 18)\%$ for temporally post-selected emission. These results establish measurable two-photon interference in deterministically integrated telecom-band emitters under experimentally accessible excitation conditions. Although photon indistinguishability remains limited by dephasing and carrier-induced fluctuations, the results identify clear pathways for improvement through radiative-rate engineering, charge stabilization, and optimized excitation schemes. By combining epitaxial growth, deterministic device fabrication, and direct C-band emission, this work advances scalable semiconductor quantum light sources compatible with existing fiber-optic infrastructure and provides a foundation for their further development toward quantum networking and photonic quantum information processing.

\subsection*{Author Contributions}

MP, YB, and AK contributed equally to this work. MP designed and processed the CBG-based single-photon sources and contributed to their spectroscopic characterization, including FEM simulations, deterministic device fabrication, and µPL measurements. YB performed the epitaxial growth, wafer bonding to integrate metallic mirror, initial optimization and characterization of the quantum-dot wafer material. AK led the spectroscopic investigation and data analysis. DMF contributed to data analysis and the preparation of the initial manuscript draft. AB supported deterministic device fabrication and processing. AKS contributed to the preparation of the initial manuscript draft, data interpretation and analysis, and manuscript review. PH developed the methodology for the planar cavity fabrication, supported the epitaxial growth, initial optimization, and characterization of the quantum-dot wafer material. SR contributed to device design and FEM simulations. IL assisted with spectroscopic characterization and data analysis, contributed to the initial manuscript draft, and reviewed the manuscript. ES supervised the epitaxial growth, planar cavity fabrication and initial optimization and characterization of the quantum-dot wafer material. SR coordinated the project, supervised the spectroscopic characterization of the CBG-based single-photon sources, and revised the manuscript.

\subsection*{Acknowledgments}

This work was funded by the German Research Foundation via the projects Re2974/23-1, Re2974/33-1 and INST 131/795-1 320 FUGG, from the Federal Ministry of Research, Technology and Space (BMFTR) via the projects MultiCoreSPS (Grant No. 16KIS1819K) and QR.N (Grant No. 16KIS2193), from Berlin Quantum (BQ), the Danish National Research Foundation through NanoPhoton -- Center for Nanophotonics, grant number DNRF147, and the QPIC1550 project, funded by the EU Horizon Europe program under Grant Agreement no. 101135785. P.\,H. was supported by the Polish National Agency for Academic Exchange under the Polish Returns program.

\subsection*{Conflicts of Interest}

The authors declare no conflicts of interest.

\subsection*{Data Availability Statement}

The data that support the findings of this study are available from the corresponding author upon reasonable request.

\subsection*{Supporting Information}

Additional supporting information can be found online in the Supporting Information section.

\bibliographystyle{unsrt}
\bibliography{References}

\end{document}


\twocolumn[
\maketitle
\vspace{1em}
]
The Supporting Information is structured as follows. Section 1 presents additional cathodoluminescence (CL) measurements. Section 2 provides a detailed description of the optical characterization setup and the measurement techniques employed, including micro-photoluminescence (µPL) time-resolved PL (TRPL), Hanbury Brown and Twiss (HBT), and Hong–Ou–Mandel (HOM) measurements. Section 3 presents a comprehensive characterization of circular Bragg grating 1 (CBG), covering its polarization-resolved, power-dependent, time-resolved, and temperature-dependent emission properties. Finally, Section 3.3 examines the dependence of the extracted HOM visibility on temporal post-selection.

\section{Cathodoluminescence mapping and spatial localization of quantum dots}

\begin{figure}
    \centering
    \includegraphics[width=\linewidth]{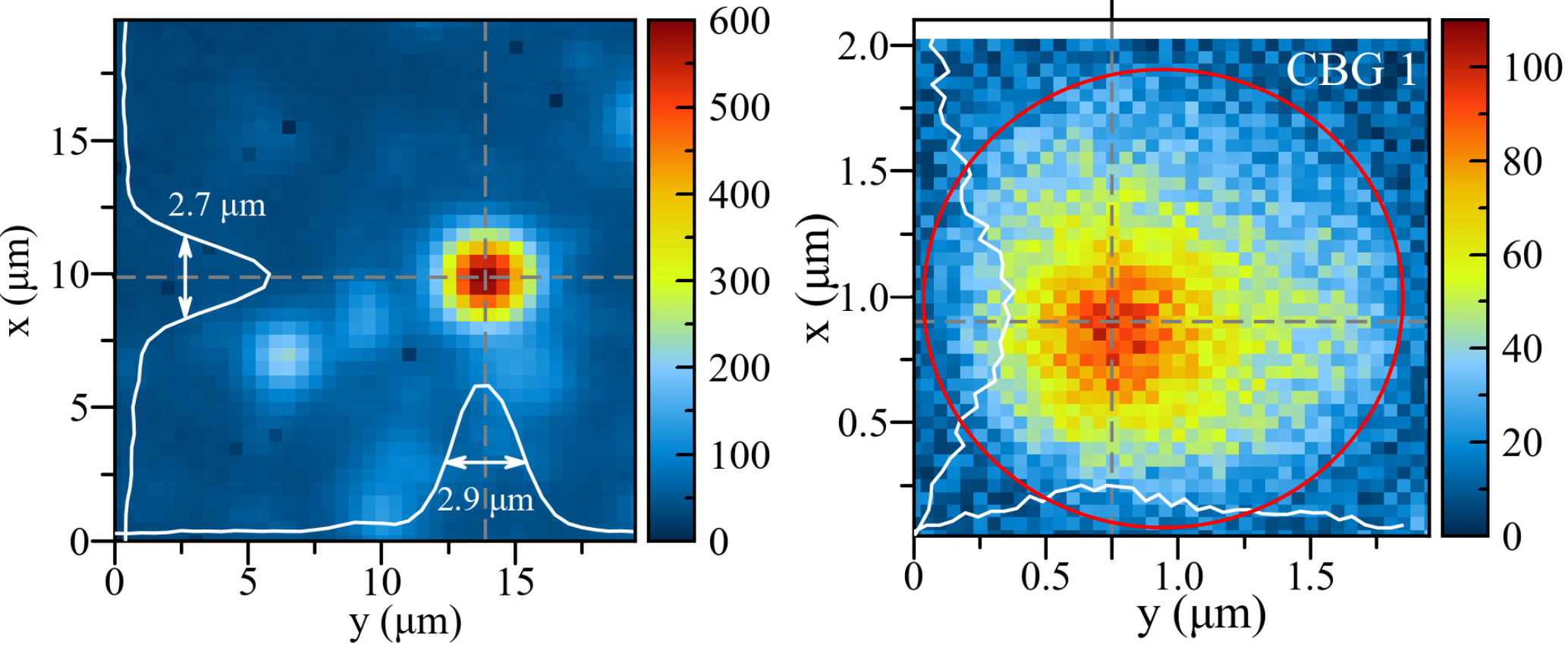}
    \renewcommand{\thefigure}{S\arabic{figure}}
    \caption{(a) CL intensity map acquired over a $20\times20~\mu\mathrm{m}^2$ area, showing multiple QD emission sites and the selected telecom C-band emitter (marked by dashed lines). Intensity profiles extracted along the dashed lines yield FWHMs of $2.7~\mu\mathrm{m}$ and $2.9~\mu\mathrm{m}$ along the ($x$)- and ($y$)-directions, respectively. (b) High-resolution CL map of the CBG1 mesa acquired over a $2\times2~\mu\mathrm{m}^2$ area. The corresponding intensity profiles exhibit FWHMs of $0.8~\mu\mathrm{m}$ and $0.9~\mu\mathrm{m}$ along the ($x$)- and ($y$)-directions, respectively, demonstrating the spatial localization of the emission within the mesa (red circle). The dashed lines indicate the positions of the extracted intensity profiles, and the color bars represent the CL intensity.} 
    \label{fig:CL_map}
\end{figure}

CL mapping was used to identify suitable C-band quantum dots (QDs) in the deterministic nanofabrication process, and to determine their spatial positions relative to the CBG structures in the post-processing analysis. Figure \ref{fig:CL_map}(a) shows a representative CL intensity map acquired over a $20\times20~\mu\mathrm{m}^2$ area, revealing multiple emitting QDs (bright spots) within the investigated sample region. The brightest QD emitting around 1540 nm was selected for device integration. Intensity profiles extracted along the orthogonal directions (dashed lines) yielded full widths at half maximum (FWHM) of $2.7~\mu\mathrm{m}$ and $2.9~\mu\mathrm{m}$ along the ($x$)- and ($y$)-directions, respectively.

To investigate the spatial localization of the selected QD with the deterministically fabricated CBG, a high-resolution CL map overlaid with an SEM image was acquired over a \(2\times2~\mu\mathrm{m}^2\) region and of CBG1, as shown in Fig. \ref{fig:CL_map}(b). The corresponding intensity profiles exhibit substantially narrower FWHMs of ($0.8~\mu\mathrm{m}$) and ($0.9~\mu\mathrm{m}$) along the (x)- and (y)-directions, respectively.

It should be noted that the measured CL emission profiles are influenced by carrier diffusion and excitation volume, therefore do not directly represent the intrinsic spatial extent of the QD. Nevertheless, fitting the spatial intensity distributions provides a reliable estimate of the emitter position, which is essential for assessing the spatial alignment between the selected QD and the fabricated CBG device.

\section{Optical setup and measurement techniques}

\begin{figure*}[!t]
    \centering
    \includegraphics[width=0.7\linewidth]{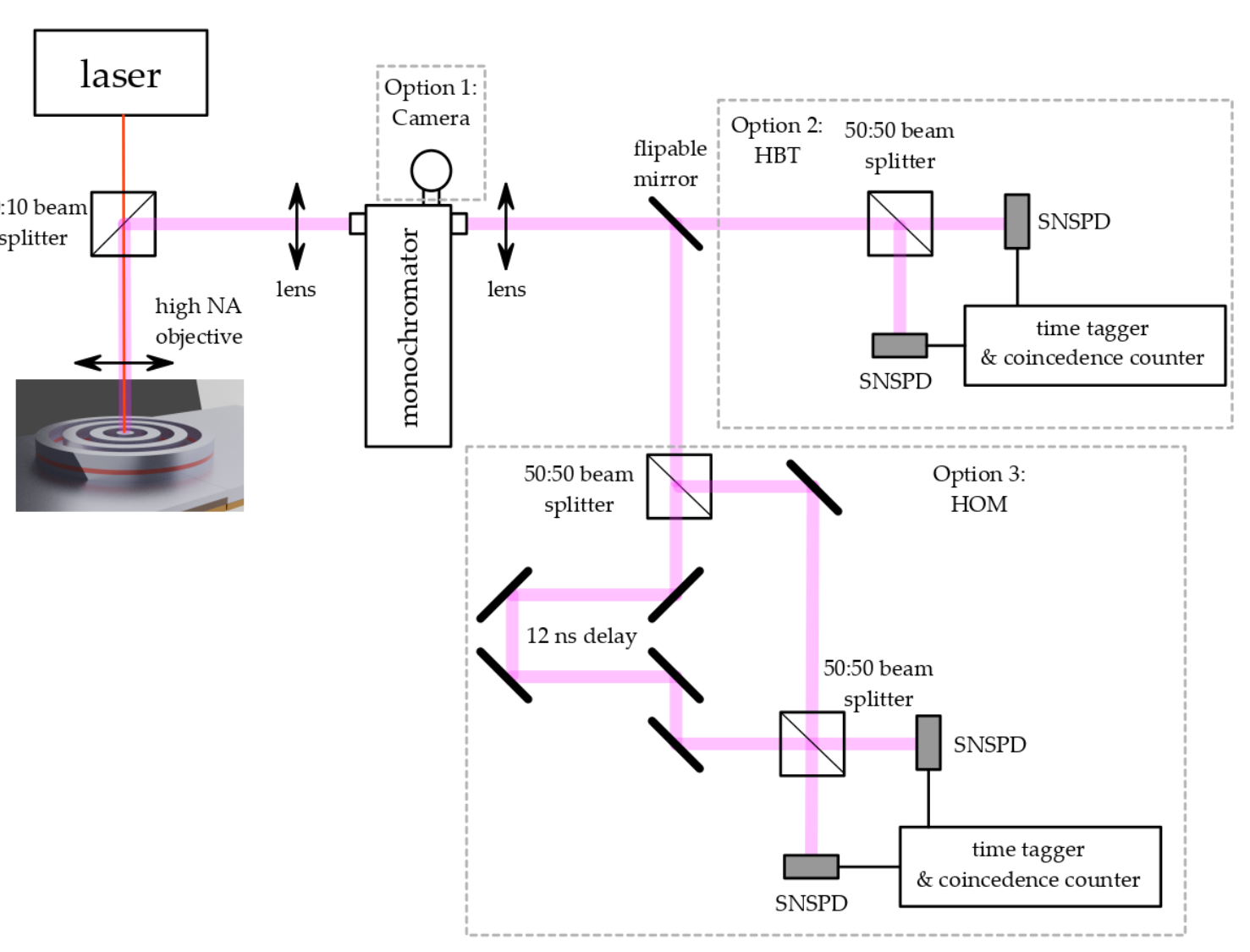}
    \renewcommand{\thefigure}{S\arabic{figure}}
    \caption{PL excited by CW or pulsed lasers is collected using a cryogenic microscope objective and spectrally dispersed by a monochromator and detected using an InGaAs 1D array detector for spectroscopic measurements (Option 1, camera) or coupled into a fiber-based optical setup for time-resolved and quantum-optical characterization. TRPL measurements correlate photon arrival times with the laser synchronization signal, while HBT measurements use a 50:50 fiber beam splitter to determine the second-order autocorrelation function. For HOM measurements, a fiber-based interferometer introduces a delay corresponding to the laser repetition period ((\qty{12.5}{\ns})), enabling interference between consecutively emitted photons. Co- and cross-polarized configurations are used to measure two-photon interference and establish the distinguishable-photon reference, respectively. All time-resolved measurements are performed using a two-channel SNSPD system connected to a quTAG time tagger.} 
    \label{fig:Setup}
\end{figure*}

Optical characterization was performed to investigate the influence of the CBG nanophotonic structures on the photon extraction efficiency (PEE) and to assess the quantum-optical properties of the emitted photons. All measurements were conducted in a closed-cycle cryostat at a sample temperature of approximately (\qty{4}{\K}). The cooling capacity of the cryostat exceeded the heat load introduced by optical excitation, ensuring stable sample temperatures throughout the experiments.

As illustrated in Fig. \ref{fig:Setup}, the µPL emitted by the QDs was collected using an apochromatic microscope objective with a numerical aperture of ($\mathrm{NA}=0.8$), optimized for the near-infrared wavelength range of approximately (\qtyrange[range-units=single,range-phrase=-]{1180}{1580}{\nm}). The collected emission was collimated and guided through a series of silver mirrors into a monochromator with a focal length of (\qty{750}{\mm}). A (\qty{50}{\micro\meter}) entrance slit defined the spatial selection of the collected signal. Depending on the grating used, the monochromator provided spectral a resolution of approximately (\qty{0.22}{\nm}) with a 150 lines/mm grating and (\qty{0.024}{\nm}) with a 900 lines/mm grating, respectively. For spectrally resolved µPL measurements, the filtered signal was directed onto a liquid-nitrogen-cooled InGaAs array detector.

The QDs were excited non-resonantly using either a continuous-wave (CW) laser at (\qty{785}{\nm}) or a pulsed laser at (\qty{800}{\nm}), with a pulse duration of (\qty{2}{\ps}). Both excitation wavelengths are above the bandgap of the surrounding semiconductor matrix, enabling carrier generation followed by phonon-induced relaxation into the QD states. Since the microscope objective was optimized for near-infrared emission around (\qty{1.5}{\micro\meter}), it was not optimized for diffraction-limited focusing of the shorter excitation wavelengths. Consequently, the excitation spot had an approximate diameter of (\qty{50}{\micro\meter}). The spatial selectivity of the measurements was therefore primarily determined by the collection path rather than the excitation spot size.

For TRPL, HBT, and HOM measurements, the monochromator served as a narrowband spectral filter, selecting the emission line of interest. The filtered PL was coupled into an SMF-28 single-mode fiber and directed towards a two-channel superconducting nanowire single-photon detector (SNSPD) system, optimized for the (\qtyrange[range-units=single,range-phrase=-]{1.3}{1.6}{\micro\meter}) wavelength range. The detectors exhibited efficiencies exceeding (\qty{75}{\percent}). Their electrical output pulses were recorded using a quTAG time tagger, which registered individual photon arrival times with respect to a common timing reference. The recorded events were subsequently used to calculate photon arrival-time correlations between the two detection channels or correlations with the laser synchronization signal.

The temporal resolution of the SNSPDs was approximately (\qty{32}{\ps}), while the timing resolution of the time tagger was below (\qty{10}{\ps}). Thus, the overall instrument response function was predominantly limited by the timing jitter of the SNSPDs.

\textbf{Time-resolved photoluminescence (TRPL)}. As illustrated in Fig. \ref{fig:Setup}, TRPL measurements were performed by directing the spectrally filtered emission into a single SNSPD channel. The photon arrival times were correlated with the synchronization signal of the pulsed excitation laser. The resulting histogram provides the temporal evolution of the PL intensity following excitation and allows the radiative decay dynamics of the QD emission to be investigated.

\textbf{Hanbury Brown and Twiss (HBT) measurements}. To characterize the single-photon emission properties, the filtered PL was directed through a fiber-based 50:50 beam splitter and distributed between the two SNSPD channels (Fig. \ref{fig:Setup}, Option 2). The time tagger recorded the photon arrival times in both channels, enabling the calculation of the second-order intensity autocorrelation function, ($g^{(2)}(\tau)$). The resulting coincidence histogram was used to evaluate the single-photon purity 1-$g^{(2)}(\tau)$ of the emission.

\textbf{Hong–Ou–Mandel (HOM) measurements}. Photon indistinguishability was investigated using the fiber-based interferometer shown in Fig. \ref{fig:Setup} (Option 3). The filtered emission was first divided into two optical paths using a 50:50 fiber beam splitter. One path was delayed by (\qty{12.5}{\ns}), corresponding to the repetition period of the pulsed excitation laser, before the two paths were recombined at a second 50:50 beam splitter. This configuration enabled the interference of consecutively emitted photons arriving simultaneously at the final beam splitter.

The polarization of the two interferometer arms was controlled to establish either co-polarized or cross-polarized configurations. In the co-polarized configuration, photons from consecutive excitation cycles were indistinguishable in polarization and could undergo two-photon interference at the second beam splitter. This interference leads to photon bunching and a suppression of coincidences between the two output channels. In contrast, the cross-polarized configuration renders the photons distinguishable and suppresses two-photon interference, providing a reference measurement for evaluating the HOM interference visibility. The coincidence histograms recorded between the two SNSPD channels were used to quantify the indistinguishability of consecutively emitted photons.

\section{Micro-photoluminescence measurements}

\begin{figure}
    \centering
    \includegraphics[width=0.8\linewidth]{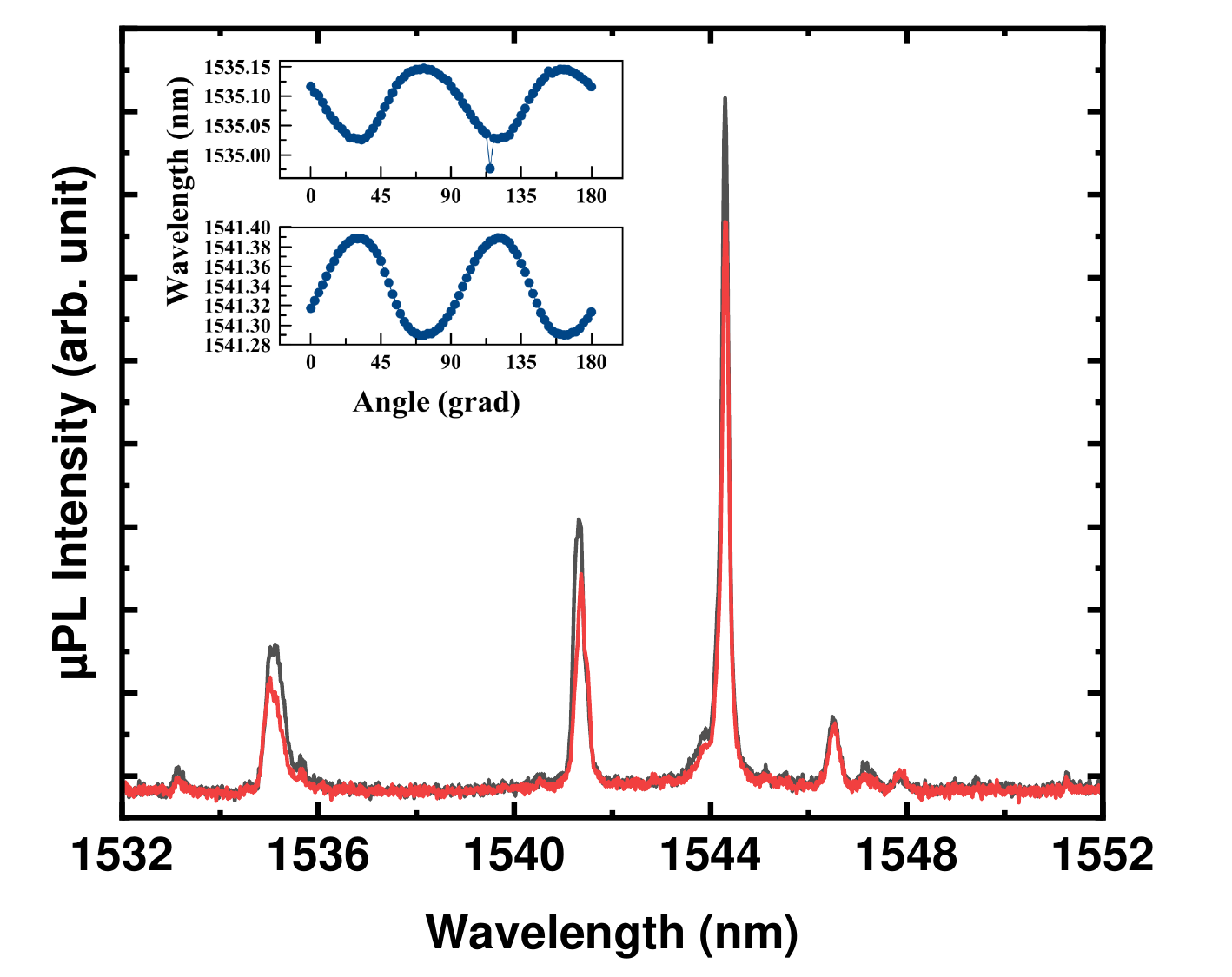}
    \renewcommand{\thefigure}{S\arabic{figure}}
    \caption{Polarization-resolved $\mu$PL spectra of CBG1 under non-resonant excitation. The red and black curves correspond to polarization angles of $0^\circ$ and $90^\circ$, respectively. Two emission lines centered at approximately 1535 and 1541 nm are identified as exciton and biecxiton, yielding a FSS of $(52 \pm 2)~\mu\mathrm{eV}$. The inset shows the polarization-dependent emission wavelengths of both transitions as a function of polarization angle.}
    \label{fig:PolSer}
\end{figure}

Polarization-resolved µPL measurements were performed to investigate the emission properties of CBG1 under non-resonant excitation. Figure \ref{fig:PolSer} shows the emission spectra recorded at polarization angles of \(0^\circ\) and \(90^\circ\), corresponding to the red and black curves, respectively. Two emission lines centered at approximately 1535 and 1541 nm are identified as exciton (X) and biexciton (XX) transition of the integrated QD. Their polarization-dependent spectral shifts were analyzed to determine a fine-structure splitting (FSS) of \((52 \pm 2)~\mu\mathrm{eV}\).

\subsection{Power-dependent emission and radiative dynamics}

\begin{figure*}
    \centering
    \includegraphics[width=0.8\linewidth]{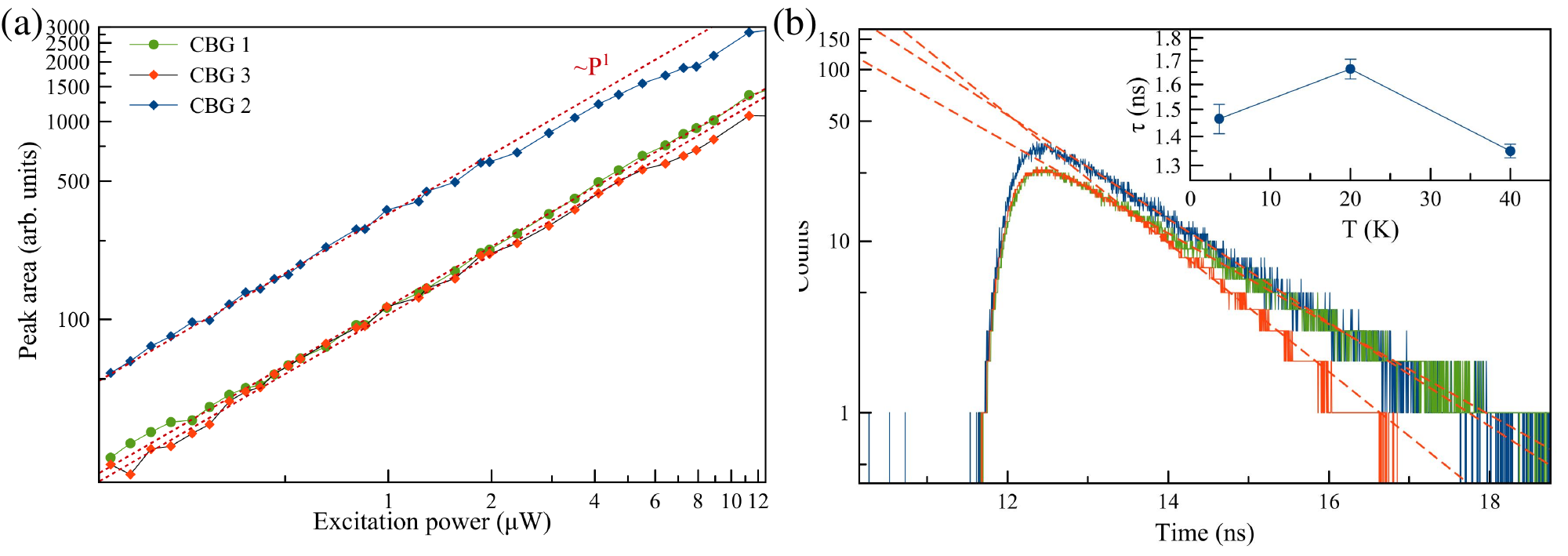}
    \renewcommand{\thefigure}{S\arabic{figure}}
    \caption{Power-dependent emission and temperature-dependent radiative dynamics of the investigated CBG structures. (a) Integrated emission intensity of the brightest emission line as a function of excitation power for CBG1, CBG2, and CBG3. All three structures exhibit an approximately linear power dependence ($I\propto P^1$), consistent with excitonic emission. (b) TRPL decay traces of the CBG1 emission line recorded at 4, 20, and 40 K. The inset summarizes the corresponding decay lifetimes of 1.45, 1.67, and 1.35 ns, respectively.}
    \label{fig:PowSer_LT}
\end{figure*}

The power dependence of the emission intensity was investigated for all three CBG-integrated QDs to characterize the nature of the observed optical transitions. As shown in Fig. \ref{fig:PowSer_LT}(a), the integrated intensity of the brightest emission line exhibits an approximately linear dependence on excitation power for CBG1, CBG2, and CBG3, following ($I\propto P^1$). This behavior is consistent with the emission originating from excitonic transitions.

The radiative dynamics were further investigated through TRPL measurements at different temperatures. Figure \ref{fig:PowSer_LT}(b) shows the decay traces of the CBG X emission line recorded at 4, 20, and 40 K. The extracted decay lifetimes are approximately (\qty{1.45}{\ns}), (\qty{1.67}{\ns}), and (\qty{1.35}{\ns}), respectively. The relatively weak temperature dependence of the measured lifetimes indicates that the radiative dynamics remain largely unchanged over the investigated temperature range.

\subsection{Temperature-dependent emission}

\begin{figure*}
    \centering
    \includegraphics[width=0.8\linewidth]{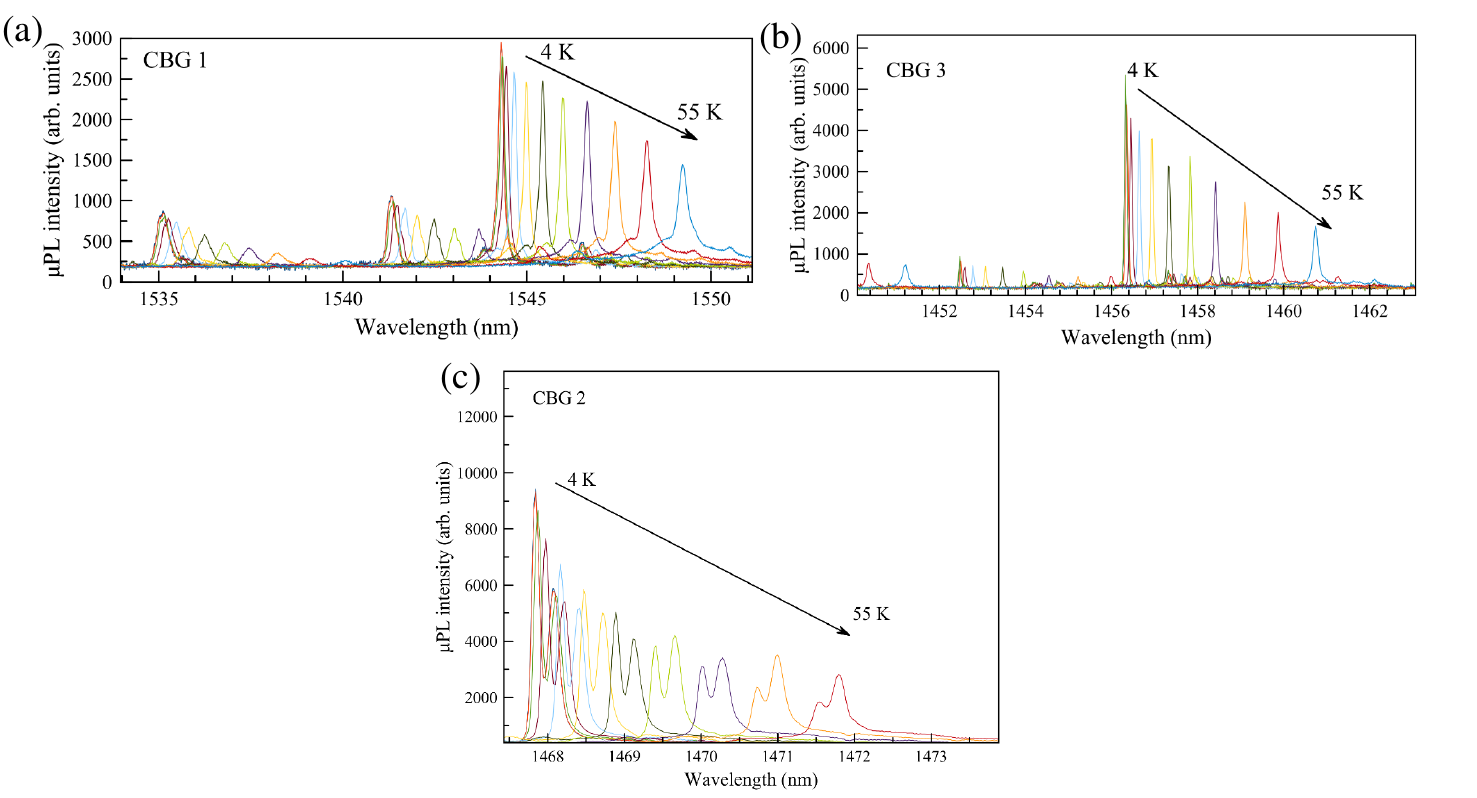}
    \renewcommand{\thefigure}{S\arabic{figure}}
    \caption{Temperature-dependent emission and thermal quenching in the CBG structures. (a--c) Temperature-dependent $\mu$PL spectra of CBG1, CBG3, and CBG2, respectively, recorded between 4 and 55 K, illustrating the thermal evolution of the emission features.} 
    \label{fig:TempSer}
\end{figure*}

The temperature-dependent emission measurements provide insight into the thermal stability of the QD–CBG devices. As shown in Fig. \ref{fig:TempSer}(a–c), increasing temperature induces a characteristic redshift and broadening of the emission spectra, consistent with bandgap renormalization and enhanced exciton–phonon interactions. The peak emission intensity also decreases due to thermally activated carrier escape from the QD into the surrounding semiconductor matrix. The extent of this thermal quenching varies among individual emission lines, reflecting differences in carrier confinement and localization. In particular, the weaker confinement of electrons compared with heavy holes makes them more susceptible to thermal activation. Despite these temperature-induced effects, the emission remains observable at elevated temperatures, highlighting the thermal robustness of the QD–CBG devices. A detailed analysis of the thermally activated processes and carrier confinement is provided in Ref. \cite{Holewa2020}.

\subsection{Dependence of HOM visibility on temporal postselection}

\begin{figure}
    \centering
    \includegraphics[width=\linewidth]{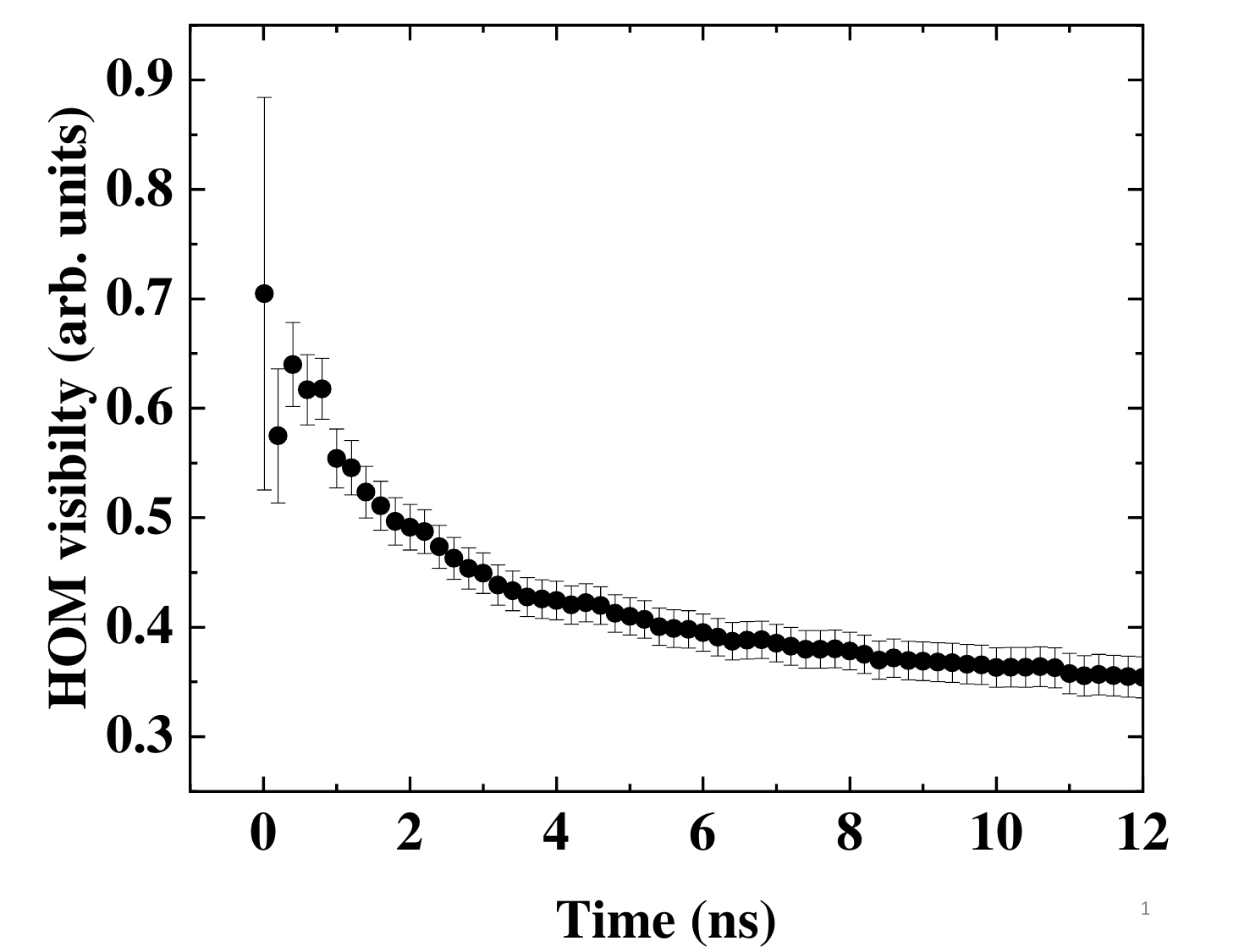}
    \renewcommand{\thefigure}{S\arabic{figure}}
    \caption{The HOM visibility was extracted from the zero-delay coincidence peak as a function of the integration time. The visibility decreases from a postselected value of \(70 \pm 18\%\) to \(35 \pm 2\%\) at the full laser repetition period of \(12.5,\mathrm{ns}\), illustrating the dependence of the measured two-photon interference visibility on the integration window.} 
    \label{fig:HOMvsTime}
\end{figure}

To investigate the dependence of the measured photon indistinguishability on the temporal postselection, the HOM visibility was evaluated for progressively increasing integration times around the zero-delay peak. As shown in Fig. \ref{fig:HOMvsTime}, the extracted visibility decreases from \((70 \pm 18)\%\) for the postselected value \((35 \pm 2)\%\) when integrating over the full laser repetition period of \(\qty{12.5}{\ns}\). This reduction indicates that the HOM visibility is influenced and limited by charge fluctuations at the ns-scale as discussed previously for short-wavelength QDs in Ref. \cite{Thoma2016}, which is typical for non-resonant excitation.

\bibliographystyle{unsrt}
\bibliography{References}